\documentclass[letterpaper,prd,showpacs,unsortedaddress]{elsarticle}[11pt]

\usepackage{graphicx}	
\usepackage{dcolumn}
\usepackage{amssymb}	
\usepackage{amsmath}	
\usepackage{hyperref}
\usepackage{multirow}
\usepackage{array}
\usepackage{bm}
\usepackage{latexsym}
\usepackage{geometry}
\usepackage{upgreek}
\usepackage{txfonts}
\usepackage{color}
\newcommand{\be}{\begin{equation}}
\newcommand{\ee}{\end{equation}}
\newcommand{\bea}{\begin{eqnarray}}
\newcommand{\eea}{\end{eqnarray}}

\journal{Physica D}

\begin{document}

\begin{frontmatter}

% ----------------------- Title Page --------------------------------

\title{Fluctuation Statistics and Diffusive Properties of the 2D Triangular Lorentz Gas in the Finite-Horizon Regime}

\author{
A.~Hosseinizadeh$^{a}$
$\footnote{Corresponding author, Email: ahmad.hosseinizadeh.1@ulaval.ca}$,
J.F.~Laprise$^{b}$,
H.~Kr\"{o}ger$^{a,c}$,
G. Melkonyan$^{a,b}$,
R. Zomorrodi$^{a}$
}
\address{
$^{a}$ {\small\sl Physics Department, Laval University,
Qu\'{e}bec, QC, G1V 0A6, Canada $\footnote{present address}$} \\
$^{b}$ {\small\sl Unit\'{e} de recherche en sant\'{e} des populations, 
CHA universitaire de Qu\'{e}bec, Qu\'{e}bec, QC, G1S 4L8, Canada} \\
$^{c}$ {\small\sl Frankfurt Institute of Advanced Studies, Goethe Universit\"at Frankfurt,
60438 Frankfurt am Main, Germany}
}
%\date{\today}

% ----------------------- Abstract -----------------------------------

%
%###############################################################################
\begin{abstract}
We study chaotic behavior and diffusion in the 2D periodic Lorentz gas in the finite-horizon regime. The dynamical observable which we consider is the length of single particle's trajectories, which moves in a triangular array of rigid disks. To analyze the chaoticity of this system, we construct a matrix of the length of trajectories and perform a level spacing analysis of the spectrum of this matrix. We find that a universal behavior occurs both in level spacing distribution and spectral rigidity. In order to examine diffusion in this classical system, we investigate the variance of the length of trajectories versus number of collisions to disks. In the case where there is a finite-horizon, such a variance scales linearly with number of bounces. This shows that a normal diffusion exists and a central limit theorem is maintained in this regime.  
\end{abstract}
%###############################################################################
%

\end{frontmatter}

%\pacs{05.45.-a; 05.40.-a}

% ----------------------- Text --------------------------------------

%\twocolumngrid

%
%
%
%###############################################################################
\section{Introduction}
\begin{figure}[tp]
\includegraphics[width=65mm,height=40mm]{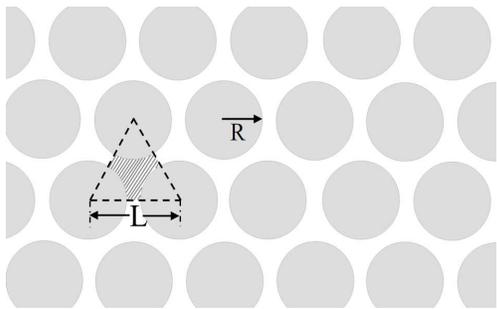}
\caption{Geometry of the 2D Lorentz gas, including a schema of the point particle trapping zone; shown between 
three adjacent disks in a triangular symmetry.}
\label{fig:Fig1}
\end{figure}

\begin{figure}[tp]
\includegraphics[width=45mm,height=40mm]{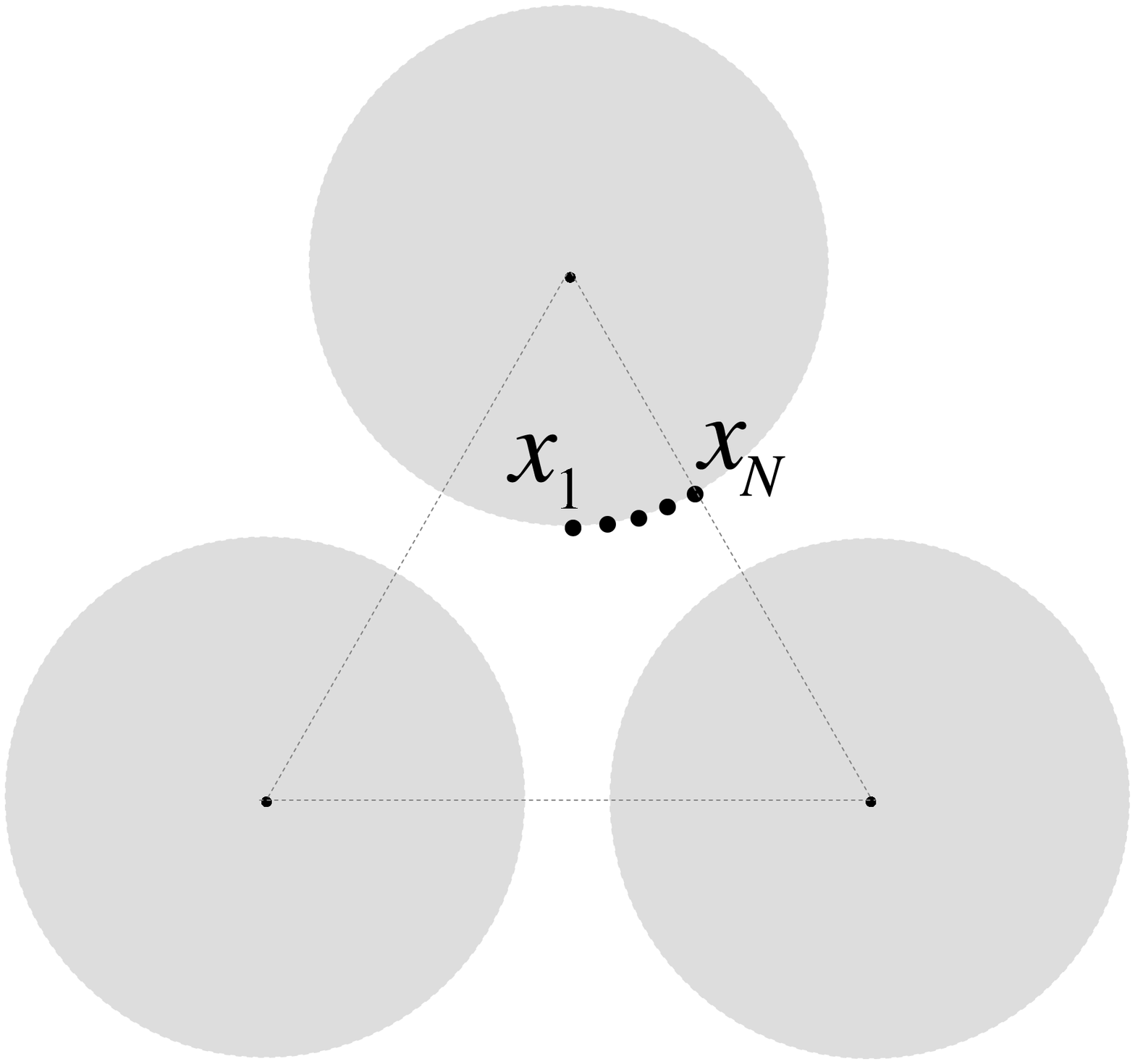}
\\
\\
\includegraphics[width=55mm,height=45mm]{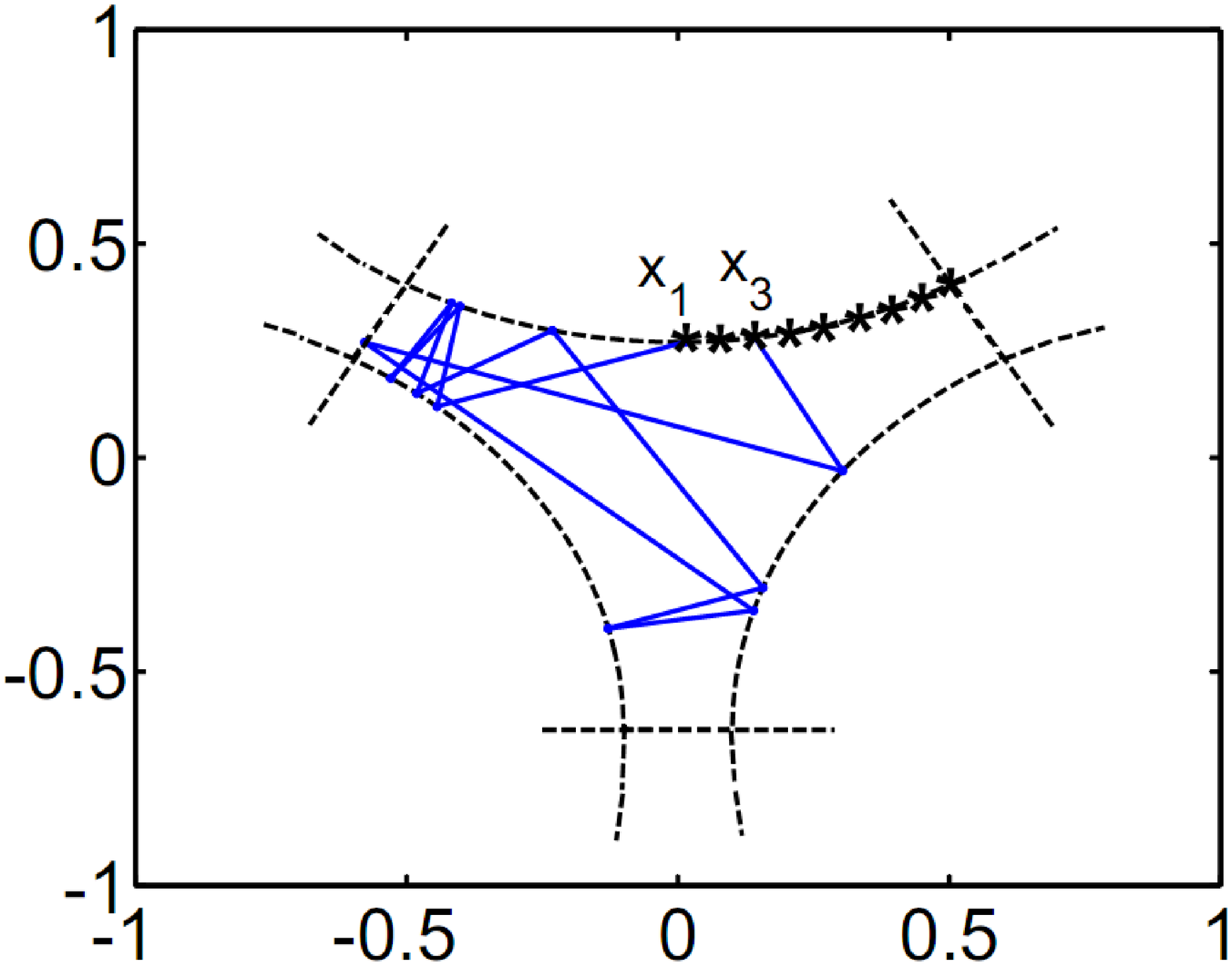}
\caption{The 2D triangular Lorentz gas in the finite-horizon regime. 
Top panel: Regular distribution of nodes $x_1,...x_N$ on the boundary of the trapping area.
Bottom panel: A sample trajectory between two boundary nodes ($x_1$ and $x_3$).}
\label{fig:Fig2}
\end{figure}

%###############################################################################
%
%
%
The 2D periodic Lorentz gas is a model equivalent to a billiard system, in which the billiard ball moves in a network of identical hard disks located on a 2D periodic regular grid~\cite{Gaspard98} (see Fig.~\ref{fig:Fig1}). The billiard ball (point-particle) alternates between free motion and collision with the disks. In the literature, Lorentz gas has been mostly considered either on a rectangular lattice (equivalent to the Sinai billiard on a torus) or on a triangular lattice. Here we consider the latter type. The dynamics of the 2D triangular Lorentz gas is basically restricted in two regimes: (i) In the finite-horizon regime ($2 < \sigma < {4}/{\sqrt{3}}$, with $\sigma=L/R$), free paths between collisions are bounded, because the scatterers are so dense, such that they block every direction of motion (note that $R$ is the radius of disks and $L$ is the central distance of two neighboring disks). (ii) In the infinite-horizon regime ($\sigma \ge {4}/{\sqrt{3}}$) there are trajectories, in which the particle can move without colliding. The rectangular Lorentz gas (Sinai billiard)~\cite{Sinai63,Sinai70} also belongs to this class. In the finite-horizon regime, the moving particle is confined to a trapping zone (see Fig.~\ref{fig:Fig2}), but in the infinite-horizon regime, it can escape to infinity. There is also a dense-packing regime with $\sigma=2$, where the particle is captured between three adjacent disks.

The distinction between the finite and the infinite regimes, defined above in geometrical terms, is also manifested in mathematical and physical laws. For example, in the finite-horizon regime, a central limit theorem (CLT) was shown to be true in the case of periodic Lorentz gas~\cite{Bunimovich81,Bunimovich91}. Chernov and Markarian~\cite{Chernov06} proved that the distribution of the travel time - to give an example of an observable - obeys the CLT, up until the $n$-th collision ($n\to\infty$). 
Therefore, it converges to a normal distribution, in the case where the number of collisions goes to infinity. 
But when the billiard particle moves with constant speed $u$, the length of trajectory $\Lambda_{n} = u t_{n}$ displays the 
same behavior, i.e., the distribution of trajectory lengths $P(\Lambda_{n})$ tends towards a Gaussian distribution for 
$n \to \infty$. Moreover, Bunimovich and Sinai~\cite{Bunimovich81} demonstrated that the velocity auto-correlation function 
decays exponentially ($\exp[-n^{\alpha}]$), which implies that normal diffusion occurs and is accompanied by a finite 
diffusion coefficient. This phenomenon has been confirmed by Machta and Zwanzig~\cite{Machta83} using a numerical 
calculation of the diffusion constant. In addition, the system converges to Brownian motion. 
 
In the infinite-horizon regime, however, no CLT has been proven yet. Furthermore, in this case there is no convergence to 
Brownian motion~\cite{Szasz07}. Bunimovich~\cite{Bunimovich83} has conjectured that the velocity auto-correlation 
function shows an algebraic decay behavior, which suggests that a standard diffusion coefficient does not exist. 
A number of researchers have carried out numerical experiments, in order to determine the decay behavior of the 
velocity auto-correlation function. In the case of the rectangular Lorentz gas, such studies have 
been performed by Zacherl et al.~\cite{Zacherl86}, Dahlqvist and Artuso~\cite{Dahlqvist96}, Zaslavsky and Edelman~\cite{Zaslavsky97}, 
Armstead et al.~\cite{Armstead03}, and Courbage et al.~\cite{Courbage08}. These studies show that the velocity auto-correlation function appears as
\be
\langle{ \vec{\text{v}} (0)~\vec{\text{v}}(t) }\rangle \sim c  / t ~ , 
\ee
where $c$ is a constant.
This indicates that the mean square displacement (the variance in position) behaves in the following manner:
\be 
\langle X^{2}\rangle \sim c t \log t ~ , 
\label{eq:Logcorr}
\ee
which was also proven by Bleher~\cite{Bleher92}. The same results were obtained for the triangular Lorentz 
gas by Friedman and Martin, Jr.~\cite{Friedman88}, as well as by Matsuoka and Martin, Jr.~\cite{Matsuoka97}.
Another chaotic system, in which the mean square displacement displays a logarithmic correction, is  the Bunimovich stadium billiard~\cite{Bunimovich74}. B\'{a}lint and Gou\"{e}zel~\cite{Balint06} showed that the distribution of the length of trajectories tends towards a normal distribution, where the limit of infinite number of collisions is concerned. In a numerical simulation (not presented here)~\cite{Laprise11}, we have shown that in the Bunimovich billiard, the variance of length $\Lambda$ obeys
\be
\mathbb{V}\text{ar}(\Lambda) \propto n \log n ~.
\label{logarithmic_variance}
\ee
One finds that the scaling factor is $ n \log n$, instead of the standard expression $\sqrt{n}$ (see Ref.~\cite{Balint06} for more details). As consequence, there is however no standard diffusion coefficient and no convergence resulting in Brownian motion in this billiard.

In addition to transport properties such as diffusion, the chaoticity of Lorentz gas has also been studied both in classical and quantum chaos~\cite{Gaspard98,Sinai70,Bohigas84,Cheon89,Primack95}. In particular, it has been shown that in a quantum Sinai billiard, statistical properties of eigenvalues display universality. This, in turn, is compatible with GOE\footnote{Gaussian Orthogonal Ensemble} predictions of random matrix theory (RMT)~\cite{Bohigas84}. But in this work, we are interested in studying this subject from a classical point of view. The strategy which we use here, has been demonstrated by Laprise et al. in Refs.~\cite{Laprise08} and~\cite{Laprise09b}. By working on a number of classical chaotic billiards (such as Bunimovich stadium billiard, cardioid billiard and an optical stadium billiard), and using methods of quantum chaos, they have found that universal behavior also occurs in classical chaos. However, in order to use the customary techniques of level spacing analysis in quantum chaos, they have computed the eigenvalues of an action (length) matrix\footnote{When the particle moves with a constant velocity, the length of the particle's trajectory is equivalent to the corresponding action. Thus, instead of an action matrix we construct a length matrix.}. As demonstrated in next section, such a matrix is constructed from the length of classical trajectories of a point-particle, which moves in the billiard zone with a constant velocity. 

As pointed out above, fully chaotic deterministic systems such as Lorentz gas, may display behavior of a random system, like Brownian motion, diffusive character and validity of CLT's. Thus, before discussing the Lorentz gas, first we take a look at a stochastic billiard. Then we examine the chaoticity of the 2D triangular Lorentz gas in the finite-horizon regime, via level spacing analysis of length matrices. And finally, we will investigate the diffusion in this system (as a physical property), in terms of trajectory length. 
%
%
%
%###############################################################################
\section{Stochastic Billiard System}
%###############################################################################
%
%
%
The collision and diffusion of molecules in
gaseous or liquid matter can be described by Brownian motion, which can be modelled by an open random
billiard. In the limit of free path length ($\Delta x \to 0$, $\Delta t\to 0$
and $(\Delta x)^2/\Delta t \to {const}$), the model describes diffusion and is mathematically described by the following differential equation, having a probabilistic interpretation~\cite{Itzykson89,Roepstorff91},
\be
\partial_t \rho = D(\vec \nabla)^2\rho~.
\ee
Let us consider the simple model of a massive particle, e.g., an atom or a molecule, which from
time to time collides with other atoms/molecules. After each collision, the particle changes its direction and
velocity. We assume that the new direction and velocity are randomly distributed. In between collisions, the
particle moves freely, maintaining direction and velocity. The length of trajectory of the random walker, having
executed $N_\text{reb}$ collisions, is given by 
\be
\Lambda=\sum_{k=1}^{N_{\text{reb}}+1} \Delta x_k \equiv \sum_{k=1}^{N_{\text{reb}}+1}\Lambda_k~.
\ee
We look at the statistical behavior of $\Lambda$ for many trajectories, having in common the same number $N_\text{reb}$ of
collisions. For statistical purpose it is meaningful to hold $N_\text{reb}$ fixed, which means that we keep fixed the average length of trajectory
\be
\langle\Lambda\rangle = (N_{\text{reb}}+1) \langle\Delta x\rangle
\ee
and $\Lambda$ fluctuates around $\langle\Lambda\rangle$. When we plot a histogram of length $P(\Lambda)$, we obtain a Gaussian distribution. This result holds provided that we consider macroscopic times, i.e. a large number of collisions. Such a behavior can be understood from the CLT: In between collisions, any length segment $\Lambda_k=\Delta x_k$ of the straight trajectory is a random number drawn from some distribution, say $P(x)$. 
The total length $\Lambda$ then is a sum of random numbers. Under general conditions on the distribution $P(x)$,
the CLT says that the sum of random variables becomes a Gaussian in the limit of large $N_\text{reb}$. As a consequence,
the length of the path of a random walker between any given starting point $x_\text{in}$ and any end point $x_\text{fi}$ follows a Gaussian distribution, if we keep fixed the number $N_\text{reb}$ of collisions, and if $N_\text{reb}$ is sufficiently large. If we construct a length matrix $\Lambda_{ij}$ , where initial and end points are taken from a given set of points ${x_1,~...~,x_N}$, then all matrix elements follow a Gaussian distribution (same variance and mean). Because $\Lambda_{ij}$ is real and symmetric, $\Lambda$ represents a matrix from a Gaussian orthogonal ensemble (GOE). Random matrix theory predicts that the level spacing distribution of the eigenvalues from a GOE matrix follows a Wigner distribution. This is similar to the statistical approach to quantum
chaos, where the level spacing distributions of energy eigenvalues from a quantum Hamiltonian are predicted
to behave in accordance with random matrix theory (i.e. like GOE in the case of a time-reversal symmetric system).
According to the BGS\footnote{Bohigas-Giannoni-Schmit}-conjecture~\cite{{Bohigas84}}, this holds when the classical counterpart of the quantum system is fully chaotic. Although the random walker is a purely random system and not a deterministic chaotic system, such observed behavior of the classical length of the random walker (in analogy to the property of Hamiltonian matrix elements in chaotic quantum systems), led us to postulate the following hypothesis: With respect to statistical fluctuations, the length matrix may play the same role in classical chaos as the Hamilton matrix in quantum chaos. In the following we investigate this hypothesis. 

Here let us mention that the BGS-conjecture does not state that the matrix elements itself of a quantum Hamiltonian are distributed like a GOE ensemble. The conjecture rather says only that the statistical fluctuations of the eigenvalue spacings, obtained from the quantum Hamiltonian, are the same as those from a GOE ensemble, giving a Wignerian distribution. In other words, it is possible that the matrix elements of the quantum Hamiltonian are distributed quite differently from a Gaussian, but nevertheless its level spacing distribution is Wignerian. Such a situation, where the distribution of matrix elements is not GOE but the level fluctuation statistics is GOE, occurs in nuclear physics. An example is the distribution of the Hamiltonian matrix elements obtained from nuclear shell model calculations~\cite{Guhr98}. In this model, there are vanishing Hamiltonian matrix elements. This implies that the number of independent matrix elements is much smaller than in a random matrix of the same size. However, Brody et al.~\cite{Brody81} could show that the 2-body residual interaction in the shell model yields matrix elements of random character following a Gaussian distribution. In particular, they showed that spectral fluctuation properties from such ensembles with orthogonal symmetry are identical to those from GOE. This implies that GOE is meaningful to predict spectral fluctuation properties of nuclei governed by 2-body interactions, though the Hamiltonian does not follow a Gaussian distribution.

%
%
%
%###############################################################################
\section{Chaotic Behavior of the 2D Lorentz Gas from the Length Matrix}
%###############################################################################
%
%
%###############################################################################
\subsection{Level Spacing Fluctuations}
%###############################################################################
%
%
In order to study the chaotic property of the Lorentz gas in the finite-horizon regime, first we construct the trajectory length matrix $\Lambda$, like in Refs.~\cite{Laprise08} and~\cite{Laprise09b}. For this purpose we consider an array of three adjacent disks , as shown in Fig.~\ref{fig:Fig2} (due to the periodicity of the system). Then we choose a set of boundary nodes $\{x_1,x_2,...,x_N\}$, that are regularly distributed on the border of the trapping region between three disks. Because of the symmetry, the boundary nodes are located on one-sixth of the border on the convex section. 

In the next step, we find the trajectory of the point-particle, as it passes between the nodes $x_i$ and $x_j$ (Fig.~\ref{fig:Fig2}). A typical trajectory is constructed by shooting the point-particle from the boundary point $x_i$ under an initial angle between $0$ and $\pi$. It moves within the trapping area and following a zig-zag path through colliding with the boundary, it finally arrives at the node $x_j$ (see Fig.~\ref{fig:Fig2}, bottom panel). In this way, we are able to build a time-reversal invariant, real and symmetric matrix $\Lambda=[\Lambda_{ij}]_{N\times N}$, in which $\Lambda_{ij}$ is the length of trajectory connecting $x_i$ and $x_j$. In fact, depending on the initial shooting angle, numerous trajectories can be constructed for a specific number of rebounds. According to this, we find an ensemble of length matrices for a particular set of boundary nodes. In practice, and to get a better statistics, we perform a level-spacing analysis (see below) on each matrix and do a superposition on the unfolded spectra.

\begin{figure}
\includegraphics[width=95mm,height=45mm]{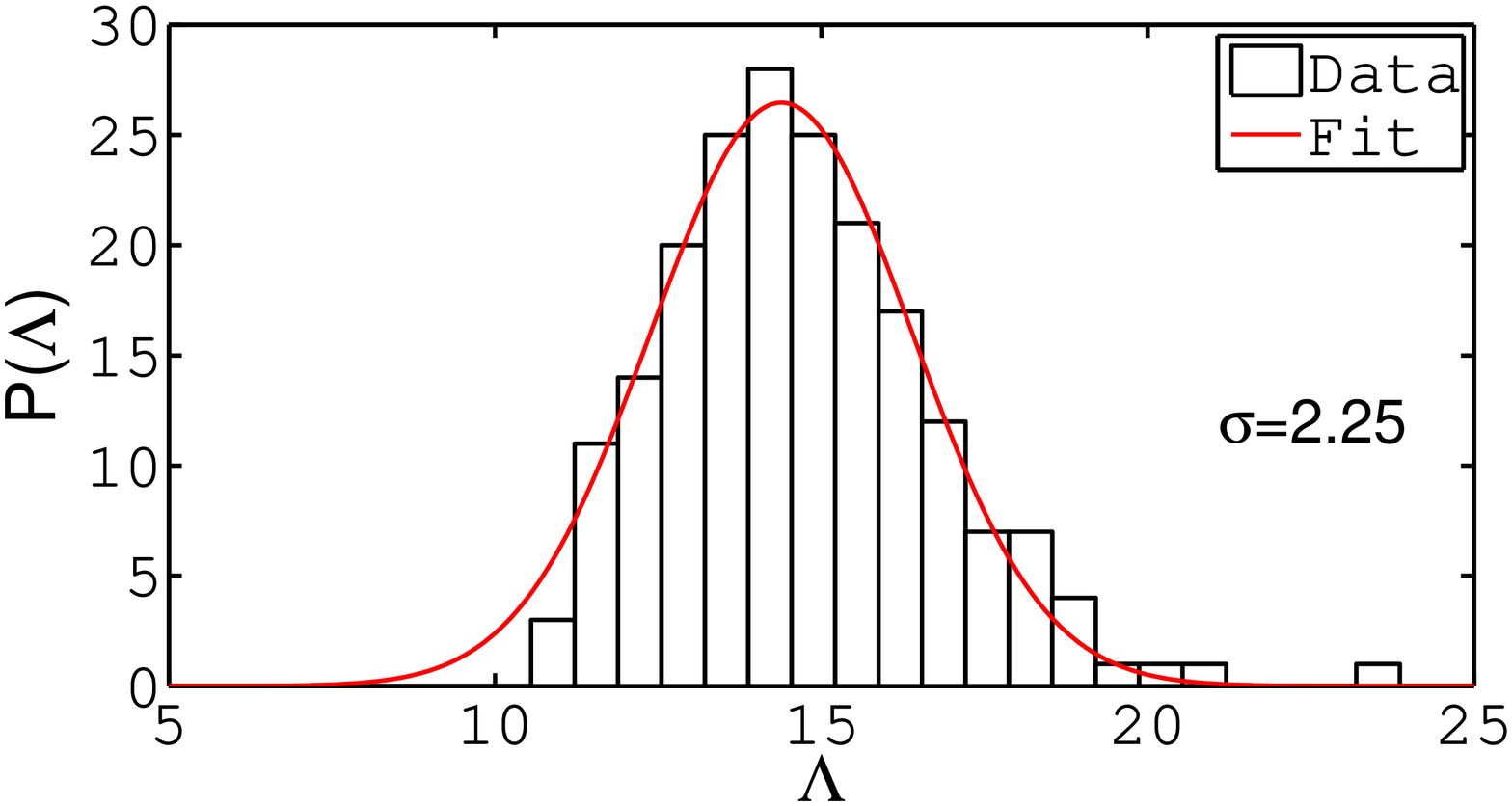}
\caption{Finite-horizon regime ($\sigma=2.25$). Distribution of the length matrix elements for $5000$ trajectories.}
\label{fig:Fig3}
\end{figure}

In a given trajectory, indeed, the consecutive segments are correlated to each other. However, two typical segments with a sufficiently large separation in terms of macroscopic travel-time are statistically independent. Therefore, such segments make an ensemble of random numbers, meaning that the point-particle follows a random walk. Through such a random walk\index{random walk}, as pointed out before, the length of a trajectory is simply given by the summation of free path lengths between successive collisions (i.e., $\Lambda=\sum_k\Lambda_k$). In reality, the so-called step lengths $\Lambda_k$ are random numbers from a Gaussian probability distribution function~\cite{{Bouchaud90}}. In the case of Lorentz gas we have also obtained the distribution of length matrix elements $P(\Lambda)$ which is near a Gaussian distribution (see Fig.~\ref{fig:Fig3}). This means that the matrix $\Lambda$ (after subtracting the mean and suitable overall normalisation) approaches a random matrix with GOE statistics~\cite{Stockmann99}. This result is generally consistent with the CLT~\cite{Bunimovich81}.
\begin{figure}
\includegraphics[width=95mm,height=45mm]{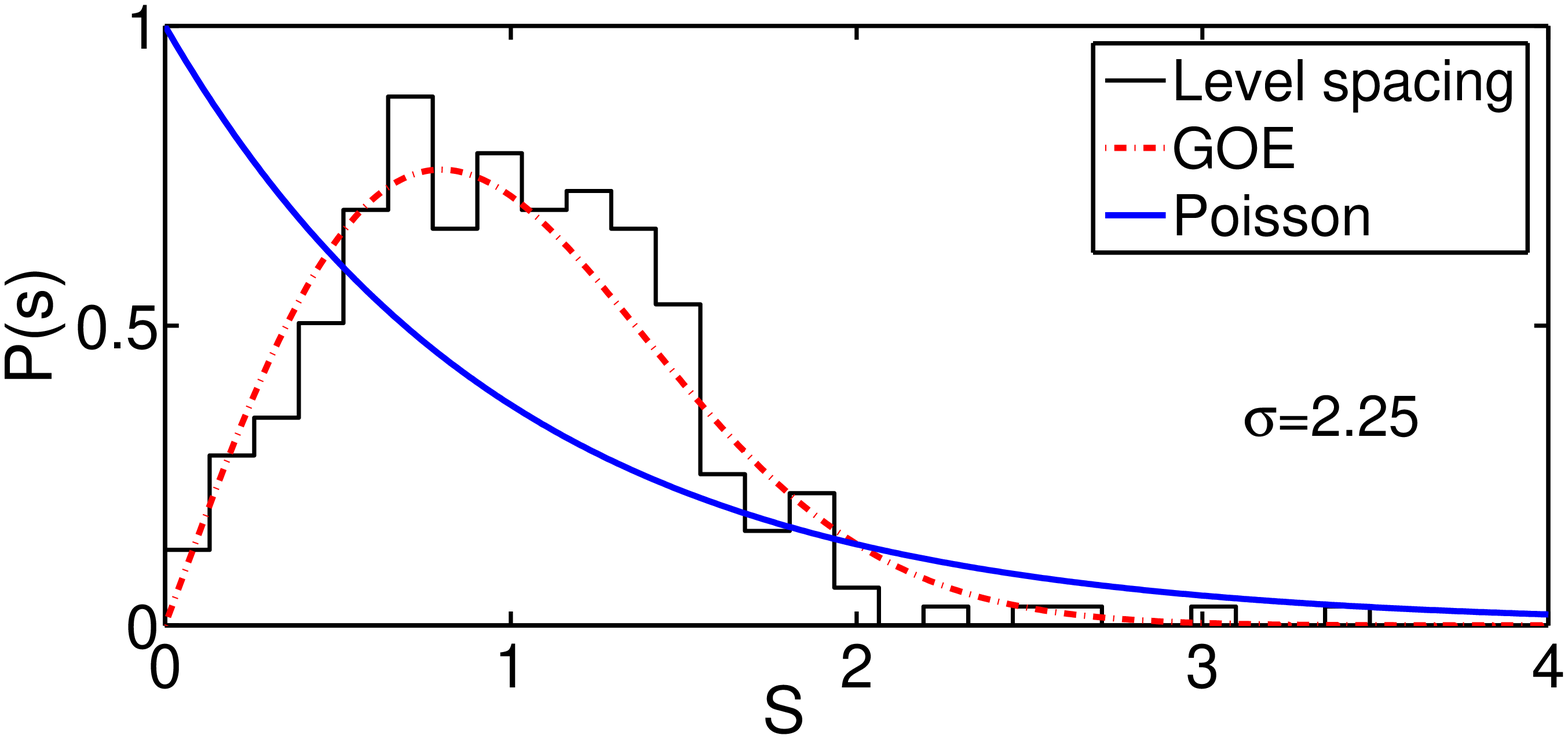}
\\
\includegraphics[width=95mm,height=45mm]{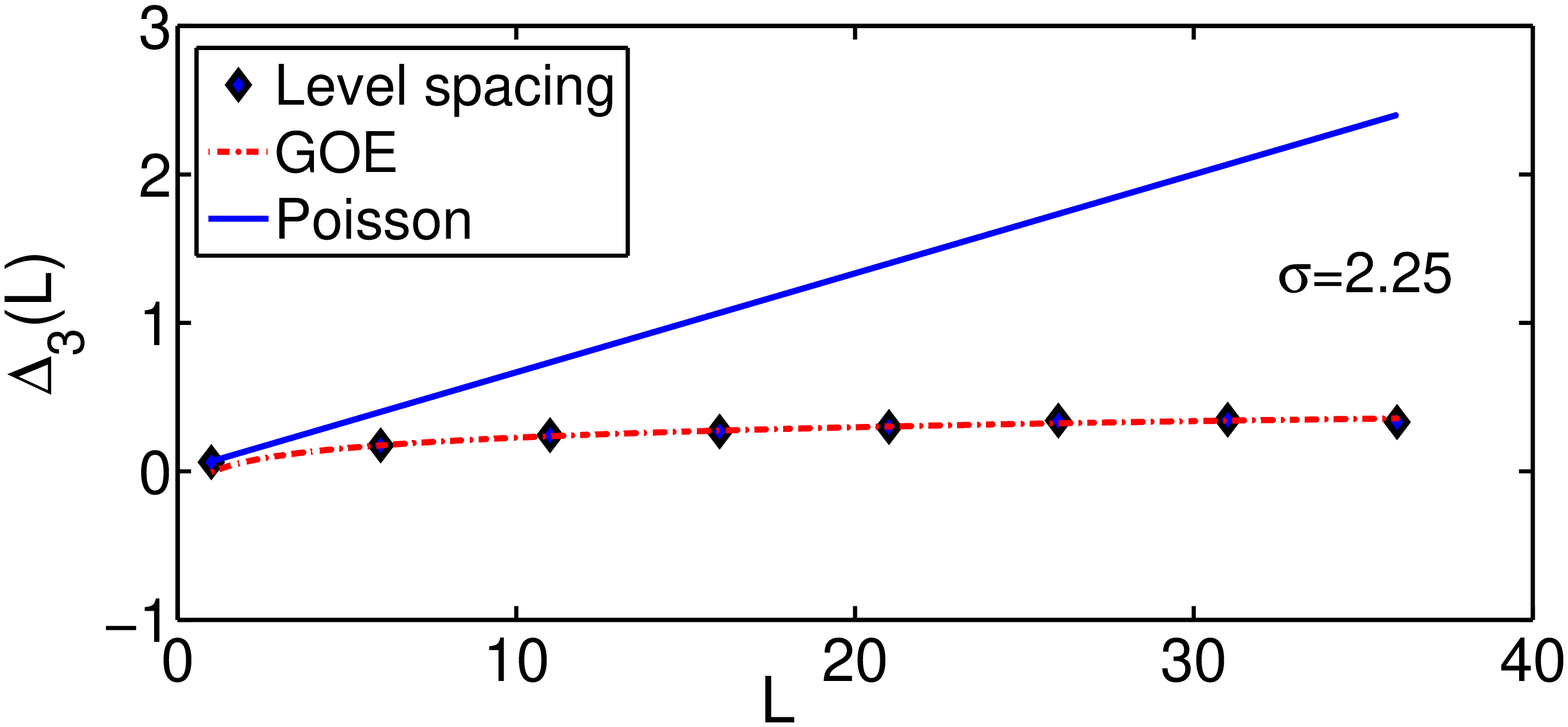}
\caption{Finite-horizon regime ($\sigma=2.25$). Top panel: Level spacing distribution $P(s)$ for the number of boundary points $N=40$ and for the number of rebounds $N_\text{reb}=15$. Bottom panel: Spectral rigidity $\Delta_3(L)$ (same parameters as the top panel). }
\label{fig:Fig4}
\end{figure}

In order to obtain the nearest-neighbor level spacing (NNS) distribution, one has to compute the spectrum of the matrix $\Lambda$ and unfold the spectrum ~\cite{Mehta91,Bohigas83}. This is a customary technique in quantum chaos where the number of energy levels tends to infinity. But, here the dimension of the length matrix is finite (it depends on the number of boundary points), and therefore, one extracts a limited number of eigenvalues. However, in quantum chaos it has been shown that a finite number of levels is also sufficient for obtaining a reliable level spacing statistics (see Ref.~\cite{Bohigas83}). Bearing this in mind, we use a Gaussian broadening~\cite{Gomez02} for the purpose of spectral unfolding of the length matrix. Then we compute the NNS distribution $P(s)$, as well as the Dyson-Mehta rigidity $\Delta_3(L)$ from the unfolded spectrum ~\cite{Bohigas75} (see the appendix for definitions). 
In the case of finite-horizon regime, as an example we choose $\sigma=2.25$ and consider a sequence of $40$ boundary 
nodes, as well as $N_\text{reb}=15$ collisions with the hard disks. Fig.~\ref{fig:Fig4} shows the NNS distribution $P(s)$ and the spectral rigidity $\Delta_3(L)$. As one observes, the distribution $P(s)$ and the spectral rigidity $\Delta_{3}(L)$ follow the GOE statistics, which is predicted by RMT~\cite{Bohigas84,Mehta91}. Similar results also hold for other values of $\sigma$ in the finite regime. This GOE property establishes universal behavior in the limit of numerous bounces for the 2D Lorentz gas with a finite-horizon. It is similar to the behavior of energy level spacing distributions in quantum chaos, in accordance with RMT.
%
%
%###############################################################################
\subsection{Evaluation of Numerical Errors}
%###############################################################################
%
%
\begin{figure}[ht]
\includegraphics[width=95mm,height=45mm]{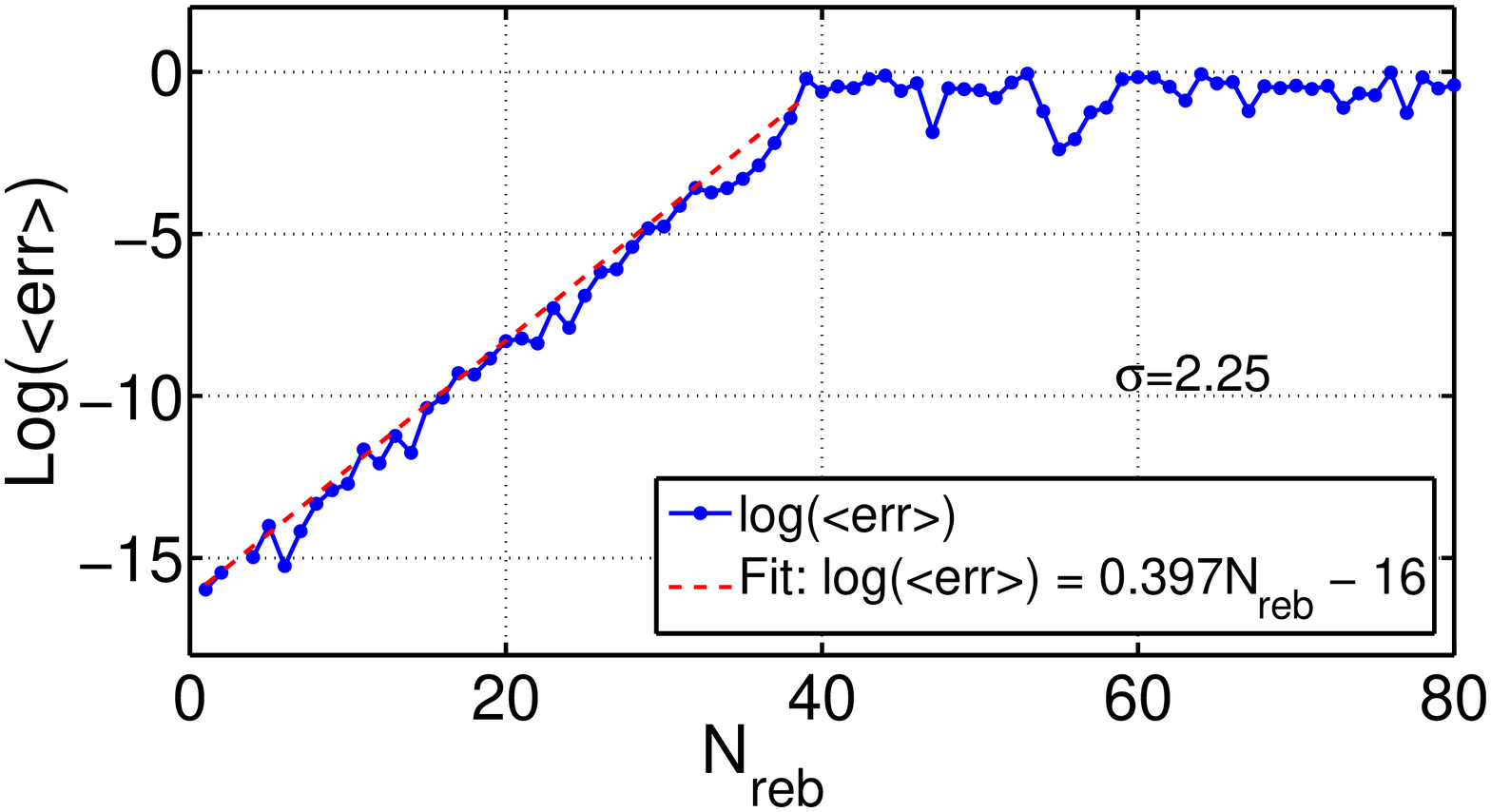}
\\
\includegraphics[width=95mm,height=45mm]{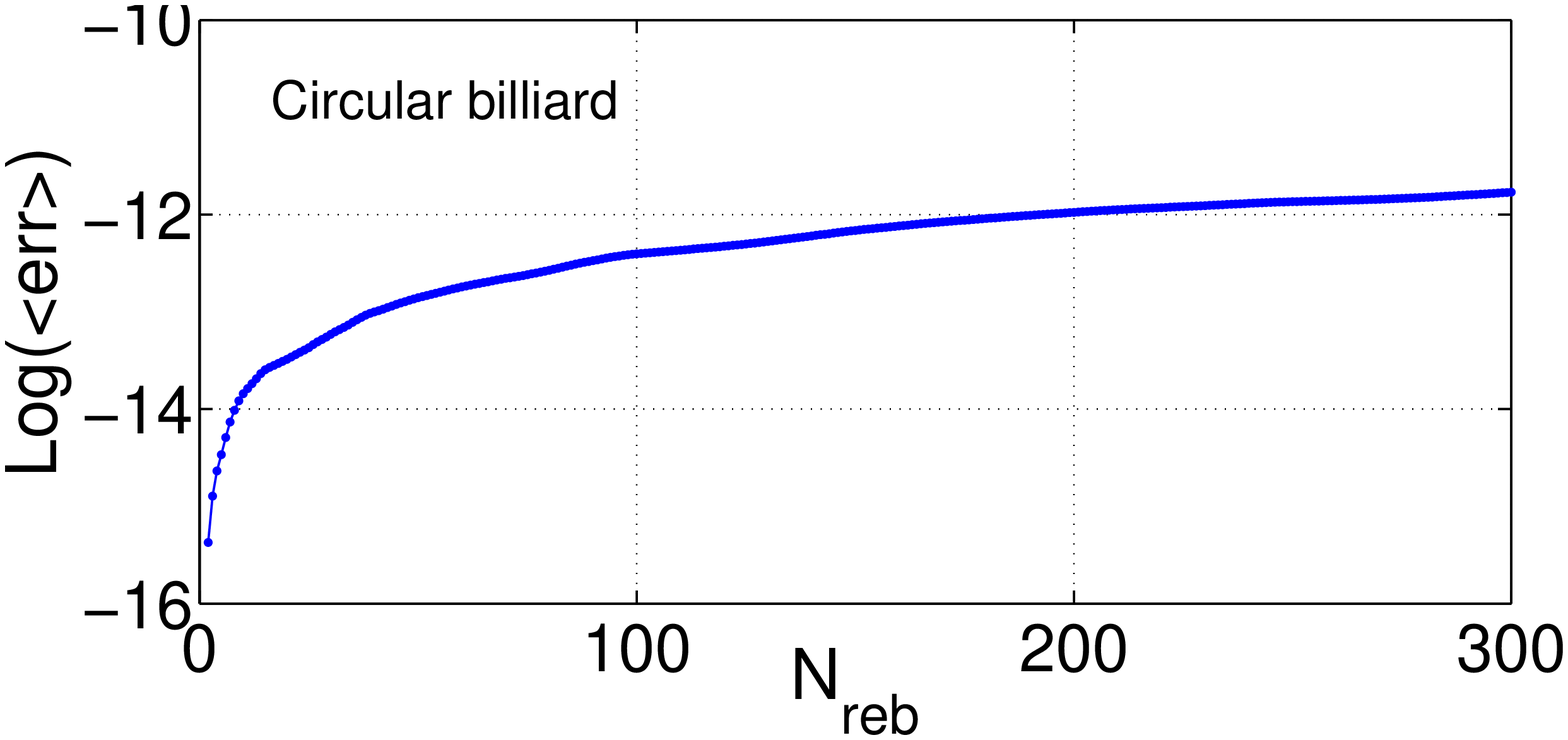}
\caption{Top panel: relative error as a function of number of rebounds in finite-horizon regime of the 2D Lorentz gas (for $\sigma =2.25$) as a fully chaotic system. Bottom panel: the same function for a circular billiard as a non-chaotic system.}
\label{fig:Fig5}
\end{figure}

In order to estimate numerical errors in the computation of a trajectory, we also compute the time-reversed trajectories
and compare the starting point of each trajectory with the endpoint of its corresponding reversed trajectory. The behavior of such an error in terms of number of rebounds is shown in Fig.~\ref{fig:Fig5} (top panel), with the following observation:
\be
\log\langle\text{err}\rangle=0.397 N_\text{reb} - 16 ~ ,~~~~~(N_\text{reb}\lesssim 40)~.
\ee
From this relation one observes that the relative error $\langle\text{err}\rangle$ increases exponentially with $N_\text{reb}$, that is $\langle\text{err}\rangle \sim \exp[0.397 N_\text{reb}]$ (as it relates to the largest positive Lyapounov exponent of the system). Figure~\ref{fig:Fig5} displays that $\log\langle\text{err}\rangle$ saturates around zero for $N_\text{reb}\approx 40$, i.e., there is a hundred percent of numerical error at this point. Furthermore, such an exponential form is not observed for integrable billiards. We have examined this for a number of integrable systems, and they display a different behavior. For example, see Fig.~\ref{fig:Fig5} (bottom panel) for a circular billiard~\cite{AhmadPhDthesis}. Hence, it can be considered as another fingerprint of chaoticity in chaotic billiards such as Lorentz gas. In our calculations we have considered trajectories with $N_\text{reb} \le 20$, which allows us to retain $8$ digits of numerical precision for the spectral fluctuation analyses. 
%
%
%###############################################################################
\section{Diffusion coefficient via the length of trajectory}
%###############################################################################
%
%
Now, let us discuss the diffusion coefficient\index{diffusion coefficient} in terms of the length of trajectory, within a trapping zone between three neighboring rigid scatterers (Fig.~\ref{fig:Fig2}). We have chosen such system, because (i) it is a fully chaotic system, having convex repellers/scatterers, (ii) it represents an example showing universality in the length matrix $\Lambda$, i.e., Wignerian form in the level spacing distribution $P(s)$, and near a Gaussian behavior in the distribution of the matrix elements $P(\Lambda)$, and (iii) its diffusion coefficient has been widely studied in the literature (for review see~\cite{Machta83,Friedman84} and references therein). 
\begin{figure}[ht]
\centering
\includegraphics[width=95mm,height=45mm]{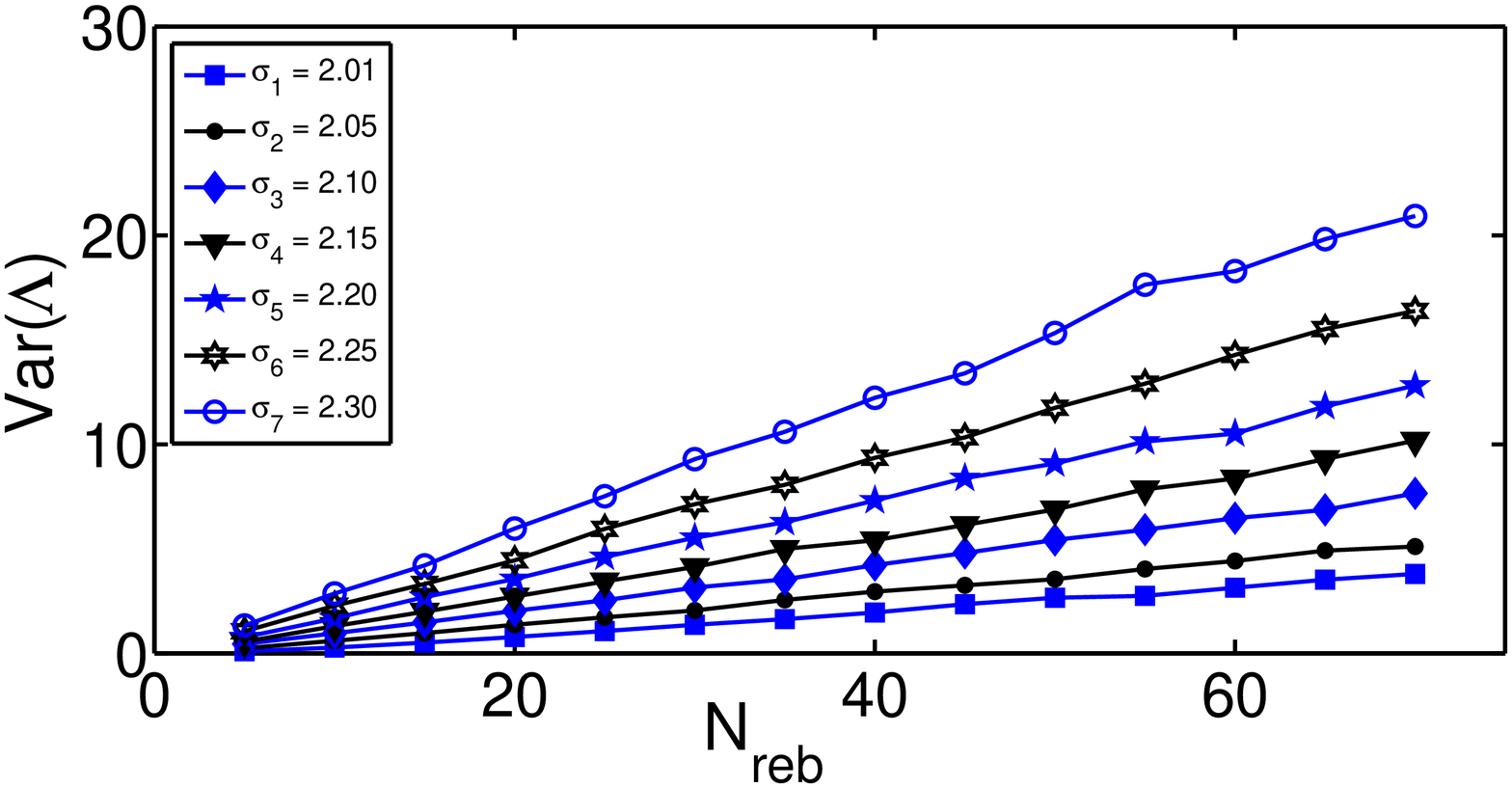} 
\\
\includegraphics[width=95mm,height=45mm]{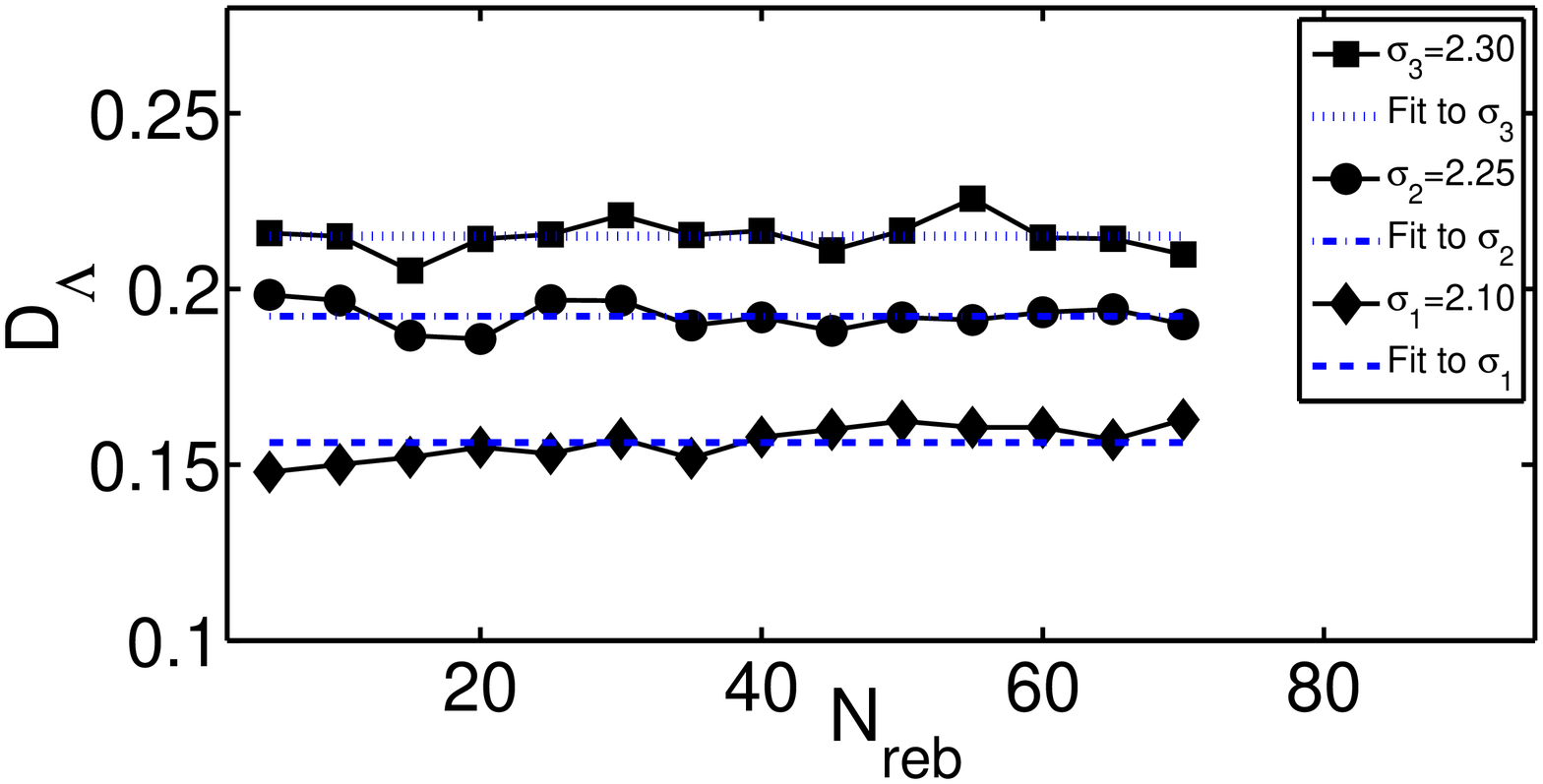}
\caption{Finite-horizon regime. Top: Variance $\mathbb{V}\text{ar}(\Lambda)$ versus number of rebounds for a variety of $\sigma$ values ($2.01, 2.05, 2.10, 2.15, 2.20, 2.25, 2.30$). Bottom: Diffusion coefficient versus number of collisions for $\sigma=2.10,~2.25,~2.30$.}
\label{fig:Fig6}
\end{figure}

As pointed out previously, for the Lorentz gas model in the finite-horizon regime, a number of properties have been proven analytically: A central limit theorem holds, the system has the character of Brownian motion (in the limit of many bounces), and a diffusion coefficient exists. The diffusive character is manifested by a linear relation between the travel time of the point-particle and the variance of position. The diffusion coefficient $\mathcal D$ (in 2 dimensions) is given by the Einstein relation
\be
\mathcal D = \frac{1}{4\tau} \Delta\vec X^2~, 
\ee
where $\tau$ denotes the time of travel and $\Delta\vec X^2$ denotes the variance of the position (see Ref.~\cite{Bleher92}). Now we will show that the diffusive character is also apparent via the length matrix which displays universal behavior. In particular, we claim that there is a linear relation between the time of travel and the variance of the length of trajectories with $n$ collisions. That is, $\Delta\Lambda^2 \propto \tau$, where travel time $\tau$ is related to mean trajectory length
of $n$ collisions $\left<\Lambda_n \right>$ and velocity $u$ via $\tau = \left<\Lambda_n \right> /u$. We define the diffusion coefficient with respect to the variable $\Lambda$ by 
\be
\mathcal D_\Lambda = \frac{\mathbb{V}\text{ar}(\Lambda)}{N_\text{reb}}~
\label{Diff_Coeff_Lambda}
\ee
(note $\mathbb{V}\text{ar}(\Lambda)\equiv \Delta\Lambda^2$). In numerical simulations, this system was found to display universal behavior manifested in the spectrum of the length matrix $\Lambda$, as well as in a Gaussian behavior of the distribution $P(\Lambda)$ (Fig.~\ref{fig:Fig3}). The results for variance of length of trajectories versus $N_{\text{reb}}$ are shown in Fig.~\ref{fig:Fig6} (top panel). We performed a statistical average over a number of $N_\text{stat}=5000$ initial conditions (position and direction). One observes that $\mathbb{V}\text{ar}(\Lambda)$ scales linearly in $N_\text{reb}$ for a variety of the geometry parameter $\sigma$ in the finite-horizon, ranging from $2.01$ to $2.30$. This linearity is in a general agreement with the results of Borgonovi et al.~\cite{Borgonovi96}. An estimate of diffusion coefficient (Eq.~\ref{Diff_Coeff_Lambda}) is also shown in Fig.~\ref{fig:Fig6} (bottom panel) for three different values of $\sigma$. From this figure one observes that $\mathcal D_\Lambda$'s are relatively constant.
\begin{figure}[ht]
\centering
\includegraphics[width=85mm,height=45mm]{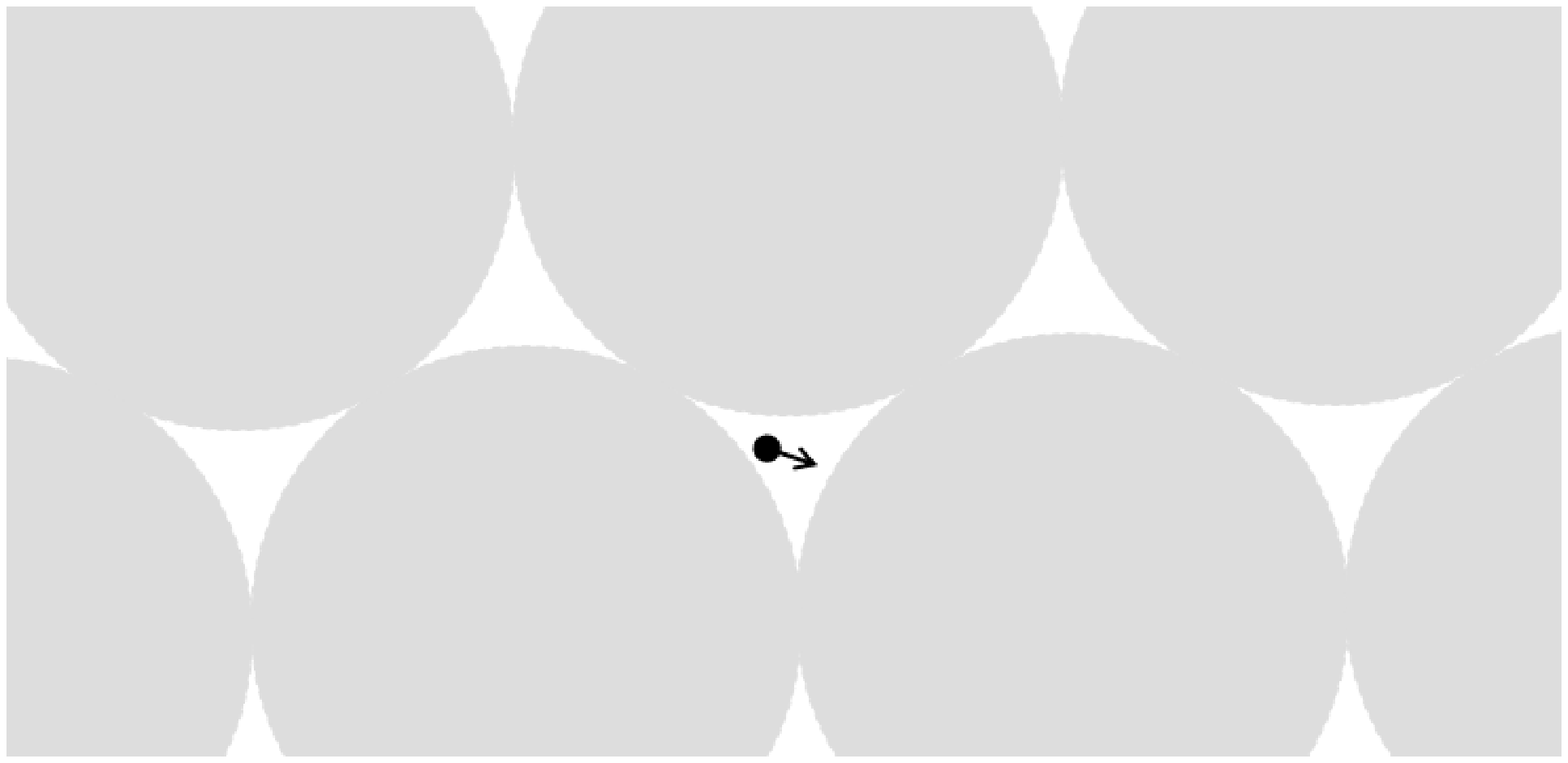} 
\\
\includegraphics[width=90mm,height=45mm]{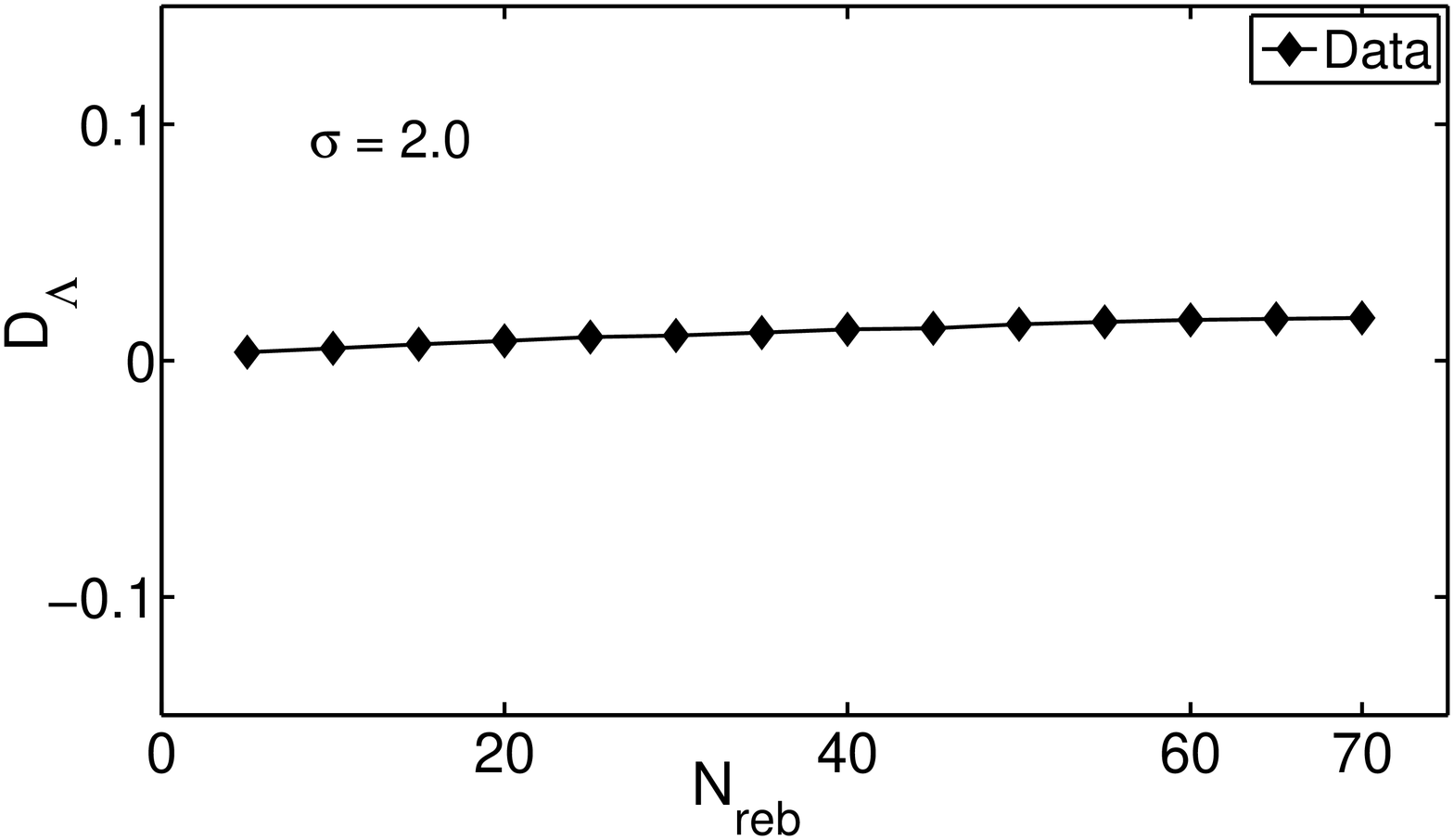} 
\caption{Top panel: The moving zone between adjacent disks in the case of the localized regime of
the 2D Lorentz gas. The point particle is trapped in this zone. Bottom panel: Diffusion constant of the Lorentz gas in the localized regime with $\sigma=2.0$.}
\label{fig:Fig7}
\end{figure}

In the limit of the dense-packing (or the so-called localized) regime (with $\sigma=2$), on the other hand, we found $\mathcal D_\Lambda \approx 0$ (Fig.~\ref{fig:Fig7}). This was expected, because the particle is caught between connected disks with no opportunity to escape beyond the trapping area. 

\begin{figure}[tp]
\centering
\includegraphics[width=95mm,height=45mm]{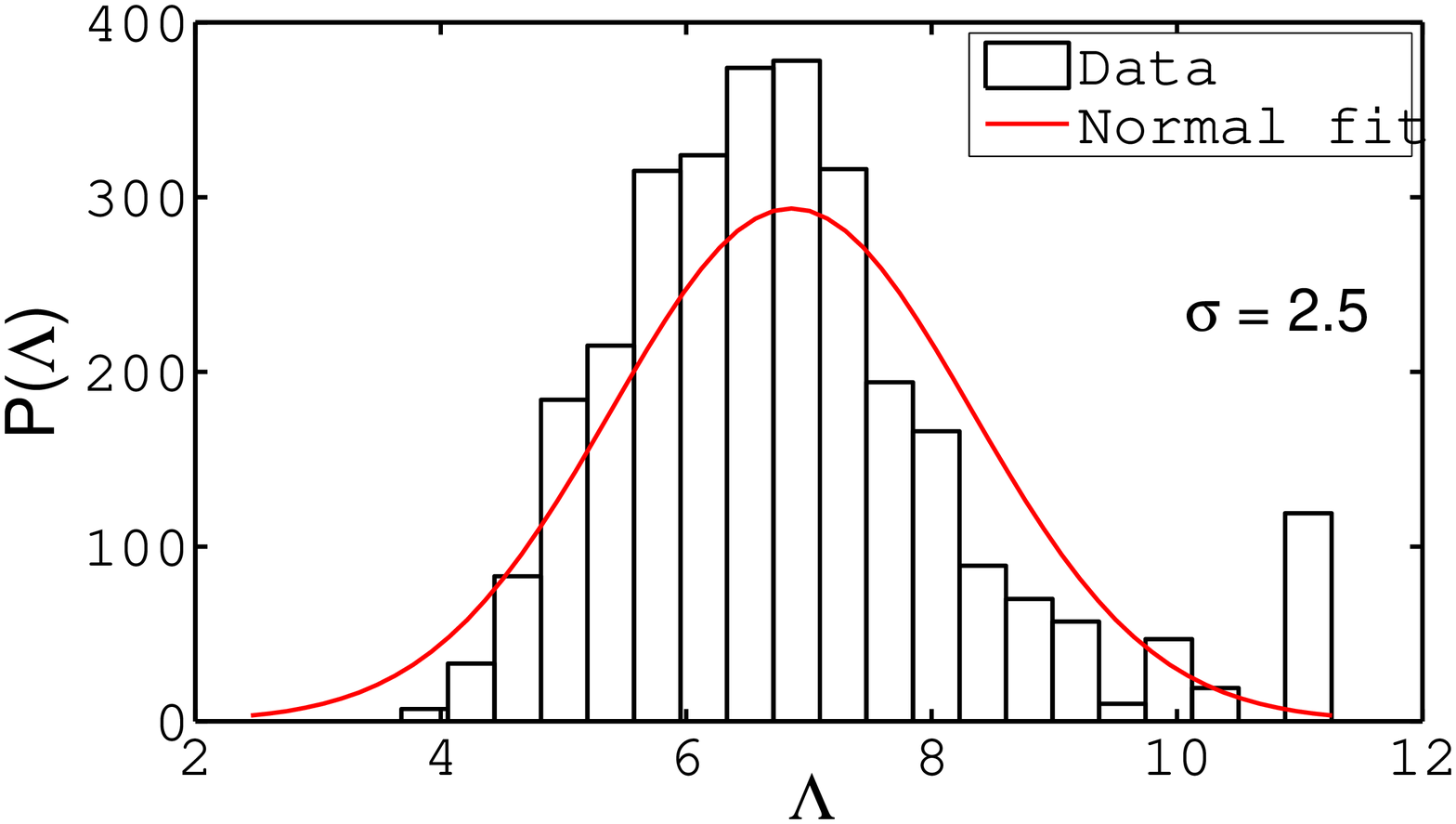} 
\caption{Distribution of length matrix elements $P(\Lambda)$ in the infinite-horizon regime with $\sigma=2.5$.
Number of rebounds and number of boundary nodes are $N_\text{reb}=15$ and $N=40$, respectively.}
\label{fig:Fig9}
\end{figure}

Finally, when going from the finite ($2 < \sigma < 4/\sqrt{3}$) to the infinite-horizon regime ($\sigma \ge 4/\sqrt{3}$), it has been predicted that the periodic Lorentz gas undergoes a change in behavior~\cite{Zacherl86,Bleher92,Friedman84,Bouchaud85}. In the infinite-horizon regime, the standard diffusion coefficient $\mathcal D$ is known to diverge~\cite{Gaspard98}. Normal diffusion goes over to anomalous diffusion, where the variance of position scales like the number of rebounds with a logarithmic correction (as mentioned in Eq.~(\ref{logarithmic_variance}))~\cite{Bleher92}
\be
\Delta\vec X^2 \propto n \log(n)~.
\ee 
We have done a numerical study of the triangular Lorentz gas in the infinite horizon regime, searching for
the scaling law of the form $\Delta\Lambda^2 \propto n \log(n)$. However, our data were too noisy, being due to long trajectories in between bounces with the disks. The reason of such trajectories is that in this regime, the point-particle finds some corridors and travels long distances without colliding to hard disks. Because of this, even the length distribution $P(\Lambda)$ differs from a Gaussian, as shown in Fig.~\ref{fig:Fig9}. With presently available computational resources we were not able to reduce numerical errors, and to obtain statistically significant results to test such a logarithmic behavior. This can be considered as a future research work.
%
%
%
%###############################################################################
\section{Summary and Remarks}
%###############################################################################
%
%
%
In this work we have studied the chaotic behavior and diffusive properties of the classical 2D triangular Lorentz gas, using the length of trajectories. Based on RMT and through level spacing analysis of length matrices, we have succeeded in finding the following results:

(i) The NNS distribution $P(s)$ and the spectral rigidity $\Delta_{3}(L)$ (which are obtained from eigenvalues of the length matrix $\Lambda$) are in conformity with a Wigner distribution. This is consistent with the GOE prediction of random matrices. 

(ii) The distribution $P(\Lambda)$ of length matrix elements resembles a Gaussian distribution. This is in agreement with the random walk and Brownian motion, as direct consequences of the CLT. In the case of infinite-horizon, the distribution $P(\Lambda)$ deviates considerably from the normal distribution. This is possibly related to the long free paths of the point-particle.  

(iii) The numerical errors (in the finite-horizon regime) exponentially evolve as a function of the number of rebounds. Such a behavior does not exist in integrable billiards, and therefore it can be considered as another signature of chaos in the system in question.
 
(iv) The variance of the length of trajectories linearly varies in terms of the number of collisions, which yields a normal diffusion coefficient when finite-horizon is involved. In the localized regime, there is no diffusion because the point-particle cannot move between rigid disks. 

Finally, we should notify that the works establishing universality in semi-classical systems are usually based on classical periodic orbits (for example, see Ref.~\cite{Argaman93}). But, here we showed that universality can also be observed using non-periodic bouncing trajectories.

%
%%
%###############################################################################
{\it Acknowledgments}.
Helmut Kr\"oger has been supported by NSERC Canada.
%###############################################################################
%
%
%
%
%%%%%%%%%%%%%%%%%%%%%%%%%%%%%%%%%%%%%%%%%%%%%%%%%%%
\section{Appendix: Spectral Statistical Analysis}
%%%%%%%%%%%%%%%%%%%%%%%%%%%%%%%%%%%%%%%%%%%%%%%%%%%
%
Consider a quantum system with the spectrum $e_i~(i=1,2,...N)$ and the level spacing $s_i=e_{i+1}-e_i$ between subsequent levels. In general, the spectrum distribution $N(e)$ of a Hermitian matrix fluctuates with its eigenvalues. One can define a mean level distribution $\overline N(e)$, and therefore, the primordial level density $N(e)$ can be splitted into an average part, as well as a fluctuating part:
\be
N(e) = \overline N(e) + N_\text{fluc}(e)~.
\ee
The average level distribution is given by
\be
\overline N({e})=\int_0^e d e'~\overline \rho (e')~,
\ee
where $\overline\rho (e)$ is the mean level spacing density function.

In fact, $\overline N(e)$ is the dominant system dependent part, and $N_\text{fluc}(e)$ exhibits the universal part of level spacing distribution. One performs such a separation because the interesting part of the level spacing distribution is the universal fluctuating part, which takes a small contribution of the distribution. In a fully chaotic system, one expects that this part is independent of the particular system, and exhibits universal fluctuations~\cite{Bohigas83}.

The main goal here is to study the sub-dominant fluctuating part of the spectrum by RMT. For this purpose, first one needs a constant mean level spacing $s$ of eigenvalues. The technique that people use in quantum chaos is to renormalize the eigenvalues, in such a way that their average separation density $\overline\rho (e)$ becomes unity. After this so-called unfolding procedure, the leading smooth part $\overline N(e)$ gets a unit spacing on statistical average. The unfolded levels are characterized by~\cite{Gomez02}
\be
\displaystyle  \{\omega_i\}=\{\overline N({e_i})\}~,~~~i=1,2,...N~,
\ee
where $\omega_i$ are dimensionless levels with the separation $s_i=\omega_{i+1}-\omega_i$, and with the density $\overline\rho (\omega)=1$. For the unfolding process one can use a Gaussian broadening method, where the average level density function $\overline\rho(e)$ is given by \cite{Gomez02}
\begin{equation}
\displaystyle \overline\rho_{\tiny G}(e)=\frac{1}{\sigma\sqrt{2\pi}}\sum_k \text{exp}\left[-\frac{(e-e_k)^2}{2\sigma^2}\right]~.
\end{equation}
From the unfolded spectrum, one computes the NNS distribution $P(s)$, which denotes the histogram of finding a separation $s$ between neighboring levels. For a chaotic system it is a Gaussian, i.e., $P_\text{Wigner}(s) =  \pi s/2~ \text{exp} [ -\pi s^2\big/4 ]~,~~s\ge 0$~\cite{AbulMagd99}. 

When the rescaled sequence of levels is available (i.e. $\{\omega_i~|~i=1,2,...,N\}$), then one can also evaluate the least square deviation of the spectral step function $N(\omega)$, from the best straight line (i.e., $a\omega+b$) fitting it in an interval $[\alpha,\alpha+L]$ of the length $L$. It is called the Dyson-Mehta rigidity (or spectral rigidity), given by \cite{Bohigas83}
\begin{equation}
\displaystyle \Delta_3(L;\alpha) = \frac{1}{L}~Min_{a,b}~\int_{\alpha}^{\alpha+L}
\left[N(\omega)-a\omega-b\right]^2~\text{d}\omega~.
\label{Delta3_function1}
\end{equation}
By replacing the functional form of $N(\omega)$ in Eq.(\ref{Delta3_function1}), one obtains a formula as~\cite{Bohigas75}
\begin{equation}
\begin{array}{ll}
\displaystyle \Delta_3(L;\alpha)=\frac{n^2}{16}-\frac{1}{L^2}\left[\sum_{i=1}^n \tilde \omega_i \right]^2 +
\frac{3n}{2L^2}\left[\sum_{i=1}^n {\tilde \omega _i}^2\right]
\\\\
\hspace{1cm}\displaystyle-\frac{3}{L^4}\left[\sum_{i=1}^n {\tilde \omega _i}^2\right]^2 +
\frac{1}{L}\left[\sum_{i=1}^n (n-2i+1)\tilde \omega_i\right] ~ ,
\label{Delta3_function2}
\end{array}
\end{equation}
with $\tilde \omega_i=\omega_i-(\alpha+\frac{L}{2})$, which means one takes the center of the interval as origin. Note that other measures of the spectral rigidity, i.e. $\Delta_1$ and $\Delta_2$, have been also introduced by Dyson and Mehta~\cite{Mehta91}, but $\Delta_3$ has proven to be more useful in RMT.
%
%
%###############################################################################

%
%
%%%%%%%%%%%%%%%%%%%%%%%%%%%%%%%%%%%%%%%%%%%%%%%%%%%%%%%%%%%%%%%%%%%%%%%%%%%%%%%%
\end{document}